\documentclass[aps,prl,groupedaddress]{revtex4}
\usepackage{graphicx}
\usepackage{here}
\usepackage{algorithm}
\usepackage{braket}
\usepackage[noend]{algpseudocode}
\usepackage{amsmath}
\begin{document}

\preprint{APS/123-QED}

\title{Implementation of a digitally encoded multigrid algorithm on a quantum computer}

\author{Peter Jaksch}
\email{peter.jaksch@gmail.com}

\date{\today}

\begin{abstract}
Multigrid has become a popular method for solving some of the most challenging real-world computational problems, 
such as computational fluid dynamics (CFD). The reason for this is the very good scaling properties of multigrid, which is
often linear, or close to linear, with respect to problem size. In this paper a method is presented, which can be used to implement a quantum version of the
multigrid algorithm. The method relies upon a quantum state that is maintained
in a equal superposition throughout the calculation, and where information is encoded digitally in the qubits in a way 
more similar to a classical computer. This differs from many existing 
quantum algorithms where information is encoded in the amplitudes of the 
quantum states in the superposition. At the core of the method is an algorithm for sharing information between the states in the superposition. 
An exponential speedup is provided for classes of problems where the solution vector can be compressed efficiently, and where a quantum compiler 
can reduce the quantum circuit depth efficiently.
\end{abstract}

\maketitle

\section{introduction}
For illustrative purposes this paper will focus on the solution of the 1D Poisson's equation with Dirchlet 
boundary conditions.  
\begin{equation}
\nabla^2u = f \label{poisson}
\end{equation}
The equation is solved by discretizing the computational domain into a (large) number of cells $N = 2^n$. The second derivative is discretized according to
\begin{equation}
\left. \frac{\partial^2 u}{\partial x^2}\right\vert_{x_i} \approx \frac{u_{i-1} -2 u_i + u_{i+1} }{\Delta x^2}. \label{laplacian}
\end{equation}
In discrete form the Poisson equation becomes a matrix problem $A\mathbf{u} =\mathbf{f}$. In this paper the elements of the solution vector $\mathbf{u}$
will be encoded as quantum states in an  equal superposition 
\begin{equation}
\mathbf{u} = \frac{1}{\sqrt{N}} \sum_{i=0}^{N-1} \ket{i} \ket{u_i}. \label{encode}
\end{equation}

\begin{figure}[H]
\centering
\includegraphics[width=7cm]{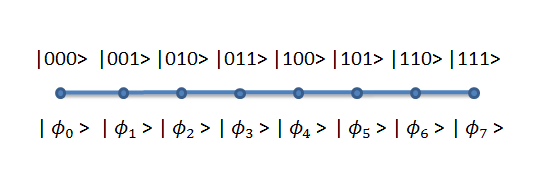}
\caption{Discretized computational domain.}
\label{multigrid}
\end{figure}

The first register contains $n$ qubits in the binary representation of the grid point number and the second register contains $k$ qubits with the data associated with that grid point. The data is encoded as in fix or floating point arithmetic. If (\ref{encode}) represents the numerical solution to equation (\ref{poisson}) a readout of the index and data registers will reveal a high resolution solution at a randomly selected 
point in the computational domain. If some particular region is of greater interest quantum amplitude amplification \cite{hoyer} can be used to to 
increase the amplitude of states in that region.

In other algorithms (see e.g. \cite{linear}) the data is encoded in the amplitudes of the quantum states. 
One problem with that approach is that it is difficult to determine high resolution data samples at specific locations in the computational domain since one would need a very large number of samples to get an accurate probabilistic reconstruction of the state. 
It is also possible to deploy highly efficient classical multigrid based methods (see \cite{lubasch}) to problems of type (\ref{poisson}). 

\section{multigrid}
The most important part of the multigrid algorithm is the V-cycle in which information is 
propagated between a hierarchical sequence of grids with different resolution.  The number of V-cycles required to
reduce the residual below a given accuracy $\epsilon$, for problems of type (\ref{poisson}), is $O(\log \epsilon^{-1})$ (see \cite{demmel}).
Assume that a starting state $\mathbf{u}_0$ has been prepared. The index register can be easily generated by applying Hadamard transformations
on each index qubit. The data register could be left in the state $\ket{0}$ or generated
using some classical function that takes the index as input. 

Next, a multigrid V-cycle (Algorithm \ref{alg1}) is applied in order to improve the starting guess.
The V-cycle relies on three operators: the restriction operator $R$ that maps from a fine grid (level $h+1$) to a coarse grid (level $h$), the interpolation (prolongation)
 operator $P$ that maps from a coarse grid to a fine grid, and the smoothing (solution) operator $S$ that improves the solution and damps high frequency errors. For the lowest level an exact solution to the discretized problem is generated. This condition can be relaxed to requiring a solution with high 
accuracy, which can be generated, e.g., by applying several iterations of the solution operator. The smoothing operator is often implemented as a Jacobi
 or Gauss-Seidel type iteration. One possible choice for the different operators in the 1D setting is
\begin{eqnarray}
R_{h+1}^h &:& u_j^{h} = \frac{1}{4}(u_{2j-1}^{h+1} + 2u_{2j}^{h+1} + u_{2j+1}^{h+1}), \quad j=1,\dots,N/2-1  \label{restriction} \\   
I _h^{h+1} &:& u^{h+1}_{2j} = u_j^{h},  \quad u^{h+1}_{2j+1} = \frac{1}{2}(u_j^{h} + u_{j+1}^{h}), \quad j = 0,\dots,N/2-1\\
S &:& u_j = \frac{1}{A_{jj}}(f_j - a_{j,j-1}u_{j-1} - a_{j,j+1}u_{j+1}) \label{smoothing}
\end{eqnarray}
Here, it has been assumed that each additional level doubles the number of grid points. 
Note that for a given point $i$ in the computational domain all three operators can be implemented as nearest neighbor actions. In other words,
if grid point $i$ has access to the data registers $\ket{\phi_{i-1}}$ and $\ket{\phi_{i+1}}$ all steps in Algorithm \ref{alg1} can be performed at this grid point.
If a Jacobi type smoother is used these operations can be performed by exploiting quantum parallelism at all grid points in $O(1)$ operations.

\begin{figure}[H]
\centering
\includegraphics[width=7cm]{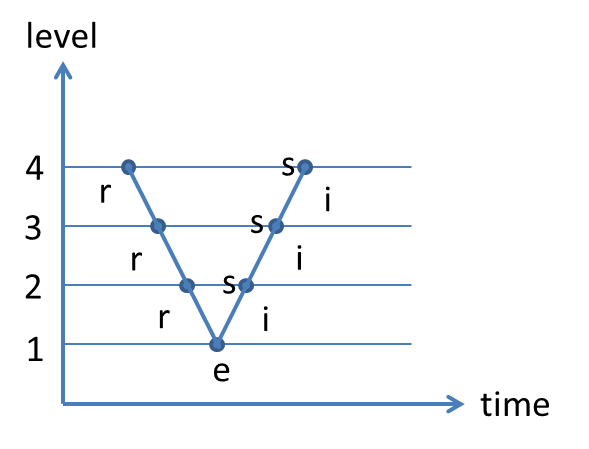}
\caption{Schematic picture of a multigrid V-cycle. r = restriction, i = interpolation, e = exact solve, s = smoothing.}
\label{multigrid}
\end{figure}

\begin{algorithm}
\caption{Multigrid V-cycle}\label{vcycle}
\begin{algorithmic}[1]
\Function{MGV}{$\mathbf{f}^{(i)}$,$\mathbf{u}^{(i)}$,i}
\If {$i == i_{min}$}
\State Solve the problem exactly
\State \Return $\mathbf{u}^{(i_{min})}$
\EndIf
\State Improve $\mathbf{u}^{(i)}$ by performing $s_0$ smoothing operations  $\mathbf{u}^{(i)} = S^{s_0}(\mathbf{f}^{(i)}$,$\mathbf{u}^{(i)})$
\State Calculate residual $\mathbf{r}^{(i)} = \mathbf{f}^{(i)} - A^{(i)}\mathbf{u}^{(i)}$
\State Restrict residual to coarser grid $\mathbf{r}^{(i-1)} = R (\mathbf{r}^{(i)})$
\State Solve $A^{(i-1)} \mathbf{e}^{(i-1)} = \mathbf{r}^{(i-1)}$ recursively on coarser grids with zero initial guess $\mathbf{e}^{(i-1)}$ = MGV($\mathbf{r}^{(i-1)},\mathbf{0},i-1$)
\State Interpolate solution to fine grid $\mathbf{e}^{(i)} = P( \mathbf{e}^{(i-1)})$
\State Correct fine grid solution $\mathbf{u}^{(i)} = \mathbf{u}^{(i)} + \mathbf{e}^{(i)}$
\State Improve $\mathbf{u}^{(i)}$ by performing $s_1$ smoothing operations  $\mathbf{u}^{(i)} = S^{s_1}(\mathbf{f}^{(i)}$,$\mathbf{u}^{(i)})$
\State \Return $\mathbf{u}^{(i)}$
\EndFunction
\end{algorithmic} \label{alg1}
\end{algorithm}

Since a very high resolution discretization is possible a simple Cartesian grid can be used. For this type of grid it is straightforward to
generate the elements of $A$ and $\mathbf{f}$ on-the-fly for many types of computational problems. 
For the boundary values an oracle is assumed that returns the boundary value based on cell index.

\newpage

\section{Quantum data register sharing}
From the previous section it is clear that the crucial part of any quantum multigrid algorithm is the sharing of information in the data registers 
between grid points in the computational domain. In this section a method is outlined for performing this task. 

\begin{algorithm}
\caption{Quantum data register sharing}\label{sharing}
\begin{algorithmic}[1]
\State Assume $U \left( \frac{1}{\sqrt{2}} \left( \ket{0} + \ket{1} \right) \right) \ket{0} \ket{0^{\otimes k}}=  \frac{1}{\sqrt{2}} \ket{0,\alpha}\ket{w_0}
+  \frac{1}{\sqrt{2} }\ket{1,\beta}\ket{w_1} $, where 
$\alpha$ and $\beta$ are 1-qubit data registers, and $\ket{w_0} $ and $\ket{w_1}$ are k-qubit working registers. 

\State Define $\ket{\Phi} = \frac{1}{\sqrt{2}}\ket{0,\alpha}\ket{w_0} + \frac{1}{\sqrt{2}}\ket{1,\beta}\ket{w_1}$.

\State Append a single qubit and perform the operation: $\ket{\Psi} = G \ket{\Phi} = \frac{1}{\sqrt{2}}\ket{0,\alpha}\ket{w_0} (\alpha \ket{0} + (1-\alpha) \ket{1}) 
+ \frac{1}{\sqrt{2}}\ket{1,\beta}\ket{w_1} (\beta \ket{0} + (1-\beta) \ket{1}) $. 
\State Rewrite $\ket{\Psi} = \sin (\pi \varphi) \ket{\chi,0}  + \cos (\pi \varphi) \ket{\xi,1}$, 
where $\sin (\pi \varphi) \ket{\chi}=\frac{1}{\sqrt{2}}(\alpha\ket{0,\alpha}\ket{w_0}  + \beta\ket{1,\beta}\ket{w_1} )$ and \newline
$\cos (\pi \varphi) \ket{\xi}=\frac{1}{\sqrt{2}}((1-\alpha)\ket{0,\alpha}\ket{w_0}  + (1-\beta)\ket{1,\beta}\ket{w_1} )$.
\State Since $\alpha$ and $\beta$ are either 0 or 1, $\sin^2 (\pi \varphi) \in \{ 0, 1/2, 1 \}$. Thus $\varphi$ only requires three bits to exactly represent all 8 solutions.
\State Let $S_{\xi}$ be a unitary operator performs a phase flip for states where the qubit in the data register is in the state 
$\ket{1}$. $S_{\Psi}=2 \ket{\Psi} \bra{\Psi} - I$ is the 
Grover diffusion operator for the state $\Psi$.  The operator $-S_{\Psi}S_{\xi}$ performs a rotation  on the two dimensional space spanned 
by $\ket{\chi,0}$ and $\ket{\xi,1}$ (see \cite{hoyer} and \cite{prakash}).
\State Write $\ket{\Psi}$ in the eigenbasis of the operator $S_{\Psi}S_{\xi}$:
$\ket{\Psi} = C_1 e^{i2 \pi \varphi}( \ket{\chi,0}  + i \ket{\xi,1}) + C_2 e^{-i2 \pi \varphi}( \ket{\chi,0}  - i \ket{\xi,1})$
\State Perform 3-bit phase estimation on $\ket{\Psi}$: 
$E\ket{\Psi} = C_1 e^{i2 \pi \varphi}( \ket{\chi,0}  + i \ket{\xi,1})\ket{\varphi} + C_2 e^{-i2 \pi \varphi}( \ket{\chi,0}  - i \ket{\xi,1})\ket{-\varphi}$. The phase estimation algorithm will suceed with probability 1 based on the previous observation that $\sin^2 (\pi \varphi) \in \{ 0, 1/2, 1 \}$. 
\State Copy the contents of the last register to a new register and switch sign conditionally on $\varphi < 0$: \newline
$CE\ket{\Psi} =  C_1 e^{i2 \pi \varphi}( \ket{\chi,0}  + i \ket{\xi,1})\ket{\varphi}\ket{\varphi} + C_2 e^{-i2 \pi \varphi}( \ket{\chi,0}  - i \ket{\xi,1})\ket{-\varphi}\ket{\varphi}$
\State Uncompute: $\ket{\Lambda} =G^{-1}E^{-1}CE\ket{\Psi} = \frac{1}{\sqrt{2}}\ket{0,\alpha}\ket{w_0} \ket{ \varphi} + \frac{1}{\sqrt{2}}\ket{1,\beta}\ket{w_1} \ket{\varphi} = \ket{\Phi}\ket{\varphi}$
\State It is possible to calculate $\beta$ from $\alpha$ and $\varphi$, and vice versa, through the relation
$\sin^2 (\pi \varphi) = \alpha^2 /2 + \beta^2 /2$. Therefore information has been shared between the data 
registers in the original superposition. 
\State Let $F$ be a quantum algorithm that utilizes the shared values.  
$\ket{\Omega} = F \ket{\Lambda} = \frac{1}{\sqrt{2}}\ket{0,\alpha,p(\alpha, \beta)}\ket{w_0} \ket{\varphi} + \frac{1}{\sqrt{2}}\ket{1,\beta,q(\alpha,\beta)}\ket{w_1} \ket{\varphi}$, where $p$ and $q$ are some functions.
\end{algorithmic} \label{alg2}
\end{algorithm}
Some comments to the above algorithm. For data registers consisting of more than a single qubit, the algorithm can simply be repeated for all qubits in the data register.   The Grover diffusion operator $S_{\Psi}$ can be implemented recursively as $S_{\Psi}=G S_{\Phi} G^{-1}$. 
In a similar fashion, $S_{\Phi}$ can be implemented as $S_{\Phi}=U(H \otimes I^{\otimes k+1})((2\ket{0}\bra{0}-I) \otimes I^{\otimes k+1})(H \otimes I^{\otimes k+1})^{-1}U^{-1}$, where $H$ is a Hadamard gate acting on the index register.
The three-qubit phase estimation part of  Algorithm \ref{alg2} requires 7 controlled applications of the Grover diffusion operator. Each Grover diffusion operator requires two applications of $U$, as already observed. Assuming that the other steps of Algorithm \ref{alg2} requires $d$ qubits in total, the resulting complexity of Algorithm \ref{alg2}  is therefore $(14 C(U) + d) \log \epsilon^{-1}$ , where $C(U)$ is the complexity of $U$ and $\epsilon$ is the accuracy of the data register.

\section{Quantum Jacobi}
In order to show how the sharing algorithm in the previous section can be used for iterative linear equation system solvers, a simple example is presented in this section.
The 1D Poisson's equation with Dirchlet boundary conditions (\ref{poisson}) is solved with the Jacobi method on a regular grid with four grid points. To make the 
situation more tangible, we may assume that the problem is to calculate the temperature in a rod with fixed cross-section when the temperature at the 
endpoints is known. In discrete form, using the discretization (\ref{laplacian}), the left hand side matrix of this  problem is 
\begin{equation*}
A = \frac{1}{\Delta x^2}
\begin{pmatrix}
-2 & 1  \\
1 & -2 & -1  \\
 & -1 & -2 & 1\\
 &  & 1 & -2
\end{pmatrix}
\end{equation*}
The right hand side vector is $(f_0, 0, 0, f_3)^{t}$, where $f_0$ ans $f_3$ are the boundary conditions (temperatures) at the endpoints. Assume that 
the starting state of the solution is a vector $(u_0^{(0)}, u_1^{(0)}, u_2^{(0)}, u_3^{(0)})^{t}$ 
A solution step of the Jacobi method for this problem consists of updating all element of the solution vector according to
\begin{equation}
u_i^{(1)} = \frac{1}{a_{ii}} \left(  f_i - \sum_{j  \neq i} a_{ij} u_j^{(0)} \right). \label{Jacobi}
\end{equation}
The quantum version of this problem starts with preparing the state 
\begin{equation}
\ket{\mathbf{u}^{(0)}} = \frac{1}{2}\ket{0,0}\ket{u_0^{(0)}}\ket{v_0^{(0)}} + \frac{1}{2}\ket{0,1}\ket{u_1^{(0)}}\ket{v_1^{(0)}} + 
\frac{1}{2}\ket{1,0}\ket{u_2^{(0)}}\ket{v_2^{(0)}} + \frac{1}{2}\ket{1,1}\ket{u_3^{(0)}}\ket{v_3^{(0)}} . \label{jacobi_start}
\end{equation}
The first register consists of two qubits and encodes the computational grid point on binary form. It is straightforward to prepare this state with a sequence of Hadamard gates. The second register is a k-qubit register that holds the data,
in this case the temperature,  encoded in fix or floating point arithmetic. At this stage, the data register will hold a starting guess (that can be efficiently prepared) to the problem. This could for example be a constant temperature, or some function that takes the index register as input and returns a temperature at that grid point location.
The third register is a working register, containing multiple qubits, possibly in a superposition.  We may assume that the preparation of (\ref{jacobi_start}) is performed with an operator $U$:
\begin{equation}
U \left( \frac{1}{2} \sum_{i=0}^{3} \ket{i} \right)  \ket{0^{\otimes k}} \ket{0^{\otimes l_0}} = \frac{1}{2} \sum_{i=0}^{3} \ket{i}\ket{u_i^{(0)}}\ket{v_i^{(0)}},
\end{equation}
where the index register has been written in decimal form. In order to implement the Jacobi step (\ref{Jacobi}) it is clear that each state (except the endpoints) 
in the superposition (\ref{jacobi_start}) needs information from its two neighboring states. In order to share information between states, Algorithm \ref{alg2} 
is used repeatedly. 
Writing  $\ket{\mathbf{u}_0} = \ket{0}\ket{\Psi_0^{(0)}} + \ket{1}\ket{\Psi_1^{(0)}}$, the operator $U (I \otimes H \otimes I^{\otimes (k + l_0)})  \left( I \otimes (2 \ket{0}\bra{0} -I) \otimes I^{\otimes (k + l_0)} \right) (I \otimes H \otimes I^{\otimes (k + l_0)})^{-1} U^{-1}$, where $H$ is a Hadamard gate acting on the last bit of the index register, implements the Grover diffusion operators $I \otimes S_{\Psi_i}^{(0)}$, $i \in \{0,1\}$ for terms 0 and 1, and 2 and 3 of $\ket{\mathbf{u}_0}$, respectively. Let $Q^{(0)}$ denote steps 1-10 in Algorithm \ref{alg2}.  
\begin{equation}
\ket{\Phi_1} = Q^{(0)} \ket{\mathbf{u}_0} =  \frac{1}{2} \left( \ket{0,0}\ket{u_0^{(0)}}\ket{v_0^{(0)}} +\ket{0,1}\ket{u_1^{(0)}}\ket{v_1^{(0)}} \right) \ket{\varphi_{0,1}^{(0)}} +  
 \frac{1}{2} \left( \ket{1,0} \ket{u_2^{(0)}}\ket{v_2^{(0)}} + \ket{1,1}\ket{u_3^{(0)}}\ket{v_3^{(0)}} \right) \ket{\varphi_{2,3}^{(0)}} . 
 \end{equation}
Calculate the shared values accoring to step 11 in Algorithm \ref{alg2} and copy the results to a new register with a quantum algorithm $F$
\begin{eqnarray}
\ket{\Phi_2} &=& F \ket{\Phi_1} =  \frac{1}{2} \left( \ket{0,0}\ket{u_0^{(0)}}\ket{u_1^{(0)}}\ket{v_0^{(0)}} +\ket{0,1}\ket{u_1^{(0)}}\ket{u_0^{(0)}}\ket{v_1^{(0)}} \right) \ket{\varphi_{0,1}^{(0)}}   \nonumber \\
 &+& \frac{1}{2} \left( \ket{1,0} \ket{u_2^{(0)}}\ket{u_3^{(0)}}\ket{v_2^{(0)}} + \ket{1,1}\ket{u_3^{(0)}}\ket{u_2^{(0)}}\ket{v_3^{(0)}} \right) \ket{\varphi_{2,3}^{(0)}} . 
\end{eqnarray}
The values $u_0^{(0)}$ and $u_1^{(0)}$ have now been shared between the first two states in the superposition. In the same way $u_2^{(0)}$ and $u_3^{(0)}$ have now been shared between the last two states in the superposition. 
Next, the values $u_1^{(0)}$ and  $u_2^{(0)}$ have to be shared between states 1 and 2 in the superposition.
Apply a cyclic (right) shift on the index register, mapping $\ket{00} \rightarrow \ket{01}, \ket{01} \rightarrow \ket{10}, \ket{10} \rightarrow \ket{11}, \ket{11} \rightarrow \ket{00}$. After regrouping terms: 
\begin{eqnarray}
\ket{\Phi_3} &=& C_r \ket{\Phi_2} =  \frac{1}{2} \ket{0} \left( \ket{0}\ket{u_3^{(0)}}\ket{u_2^{(0)}}\ket{w_3^{(0)}}  + \ket{1}\ket{u_0^{(0)}}\ket{u_1^{(0)}}\ket{w_0^{(0)}} \right) \nonumber \\
&+& \frac{1}{2} \ket{1} \left( \ket{0}\ket{u_1^{(0)}}\ket{u_0^{(0)}}\ket{w_1^{(0)}}  + \ket{1} \ket{u_2^{(0)}}\ket{u_3^{(0)}}\ket{w_2^{(0)}} \right) := \ket{0}\ket{\Psi_0^{(1)}} + \ket{1}\ket{\Psi_1^{(1)}}.
 \end{eqnarray}
 In order to simplify the notation, the states $\ket{\varphi}$ have been absorbed in the working register, increasing its length to $l_1$.
 It can be verified that the Grover diffusion operators for this state can be implemented as $I \otimes S_{\Psi_i}^{(1)} = C_r F ( I \otimes S_{\Psi_i}^{(0)} \otimes I^{\otimes l_1} ) F^{-1} C_r^{-1}$ for $i \in \{0,1\}$. 
 Share the register values with Algorithm \ref{alg2} again: 
 \begin{eqnarray}
 \ket{\Phi_4} &=& Q^{(1)}  \ket{\Phi_3} =   \frac{1}{2} \ket{0} \left( \ket{0}\ket{u_3^{(0)}}\ket{u_2^{(0)}}\ket{w_3^{(0)}}  + \ket{1}\ket{u_0^{(0)}}\ket{u_1^{(0)}}\ket{w_0^{(0)}} \right) \ket{\varphi_{0,3}^{(1)}}  \nonumber \\
 &+&  \frac{1}{2} \ket{1}  \left( \ket{0}\ket{u_1^{(0)}}\ket{u_0^{(0)}}\ket{w_1^{(0)}}  + \ket{1} \ket{u_2^{(0)}}\ket{u_3^{(0)}}\ket{w_2^{(0)}} \right) \ket{\varphi_{1,2}^{(1)}} .
 \end{eqnarray}
 By again utilizing  step 11 in Algorithm \ref{alg2} and copying the results to a new register with a quantum algorithm $G$, applying a cyclic left shift $C_l$ and absorbing garbage in the working register the resulting state is 
 \begin{eqnarray}
\ket{\Phi_5} &=& C_l G \ket{\Phi_1} =  \frac{1}{2}  \ket{0,0}\ket{u_0^{(0)}}\ket{u_1^{(0)}}\ket{u_3^{(0)}}\ket{q_0^{(0)}} +\ket{0,1}\ket{u_1^{(0)}}\ket{u_0^{(0)}}\ket{u_2^{(0)}}\ket{q_1^{(0)}}     \nonumber \\
 &+& \frac{1}{2}  \ket{1,0} \ket{u_2^{(0)}}\ket{u_3^{(0)}}\ket{u_1^{(0)}}\ket{q_2^{(0)}} + \ket{1,1}\ket{u_3^{(0)}}\ket{u_2^{(0)}}\ket{u_0^{(0)}}\ket{q_3^{(0)}} . 
\end{eqnarray}
Now, assuming that the elements of left hand side matrix $A$ and the right hand side vector $f$ can be efficiently calculated, everything is in place for the Jacobi step J. 
\begin{equation}
\ket{\mathbf{u}^{(1)}} = J \ket{\Phi_5} =  \frac{1}{2}\ket{0,0}\ket{u_0^{(1)}}\ket{v_0^{(1)}} + \frac{1}{2}\ket{0,1}\ket{u_1^{(1)}}\ket{v_1^{(1)}} + 
\frac{1}{2}\ket{1,0}\ket{u_2^{(1)}}\ket{v_2^{(1)}} + \frac{1}{2}\ket{1,1}\ket{u_3^{(1)}}\ket{v_3^{(1)}} .
\end{equation}
This is of the same form as (\ref{jacobi_start}) and the process can be repeated.

\section{Quantum multigrid}

It is obvious that the example in the previous section can be extended to registers of arbitrary size. Also, since the multigrid operators (\ref{restriction}) - (\ref{smoothing})
only rely on nearest neighbor action, the same approach can be used to implement these operators. However, in order to implement the full multi grid algorithm in an efficient way, a number of assumptions have to be introduced. Denote the exact solution vector to the problem $\phi^*$, with elements $\phi_1^*, ..., \phi_N^*$. Let $U^{(j)}$ be the operator that takes the problem from the starting
state to the state after multigrid V-cycle $j$. In order to prevent the complexity of the 
Grover diffusion operators to increase exponentially, we consider cases for which there exists an efficient approximation $\tilde{U}^{(j+1)}$, such that, 
\begin{eqnarray}
U^{(j+1)} \left( \sum c_i \ket{i} \ket{0^{\otimes k}} \ket{0^{\otimes l}} \right) &=& \sum c_i \ket{i}\ket{\phi}_i^{(j+1)} \ket{v_i}  \label{u_new} \label{super} \\
\tilde{U}^{(j+1)} \left( \sum c_i \ket{i} \ket{0^{\otimes k}} \ket{0^{\otimes l}} \right) &=& \sum \tilde{c}_i \ket{i}\ket{\tilde{\phi}_i^{(j+1)}} \ket{\tilde{v}_i}, \label{compressed}
\end{eqnarray}
where the complexity of $\tilde{U}^{(j+1)}$ is at most the complexity of $U^{(j)}$ plus a constant. For small numbers $\delta$ and $ \nu$;
the amplitudes $c_i$ are assumed to satisfy
\begin{equation}
|c_i - \tilde{c}_i| < \delta; 
\end{equation}
the states $\ket{\tilde{\phi}_i^{(j+1)}}$ are superpositions of norm 1 that satisfy 
\begin{equation}
|| \ket{\tilde{\phi}_i^{(j+1)}} - \ket{\hat{\phi}_i^{(j+1)}} ||< \nu 
\end{equation}
for some pure states $\ket{\hat{\phi}_i^{(j+1)}}$
that are close enough to $\phi_i^{(j+1)}$ in a fix or floating point arithmetic sense, according to 
\begin{equation}
\frac{|\hat{\phi}^{(j+1)} - \phi^*|}{ |\hat{\phi}^{(j)} - \phi^*|} \leq \frac{1}{k} < 1; \label{conv_rate}
\end{equation} 
$ \ket{\tilde{v}_i}$ is now a garbage register that can be an arbitrary superposition of norm 1 that does not have to be close to $\ket{v_i}$.
From equation (\ref{conv_rate}) it can be deduced that the number of multi grid V-cycles needed to reduce the residual below a given accuracy $\epsilon$, for problems of type (\ref{poisson}), changes from $O(\log \epsilon^{-1})$ to $O(\log \epsilon^{-1} / \log k)$. In order to prevent the states in the superposition (\ref{super})
to decay too much, $\delta$ and $\nu$ have to satisfy
\begin{equation}
\delta, \nu \lesssim \log k / \log \epsilon^{-1}
\end{equation}

The above assumptions  cannot be expected to hold in the general case. Therefore, it is important to identify cases where they might be applicable. Real-world 
problems usually satisfy some degree of smoothness. For reasonably well behaved problems it can be expected that the solution vector can be compressed 
with classical (lossy) signal compression algorithms such as Fourier or Wavelet transforms. This assumption is also used in \cite{lubasch}. More precisely, we will assume here that there exists a function $F$,
that can be implemented efficiently with a classical circuit, that takes an index $i$, such that 
\begin{equation}
|F(i) - \phi_i^*| < \varepsilon 
\end{equation} 
in time $O(\text{poly}( \log N))$ for all elements in the solution vector $\phi^*$. A quantum circuit $Q$ can be constructed by preparing $N$ states in an equal superposition according to equation (\ref{encode}) and then applying $F$ to each state in parallel. Obviously, $Q$ is an efficient approximation for the full quantum multi grid algorithm.
It is conjectured here that for problems that satisfy this property, also the intermediate solution vectors $\phi^{(j)}$ after each V-cycle can be approximated efficiently according to (\ref{compressed}) - (\ref{conv_rate}), with a number of quntum gates that is polynomial in $\log N$. To summarize, for the class of problems discussed 
here, there exists a quantum circuit that implements the full quantum multi grid algorithm with an exponential speedup compared to the best classical algorithm 
(classical multigrid). However, we have so far only discussed the existence of the efficient approximations  $\tilde{U}^{(j)}$ but not how to construct them. 
In other words, we know that a certain quantum circuit can be compressed but we do not know how to perform this compression.
It has to be assumed that, given their existence, a quantum compiler can construct $\tilde{U}^{(j)}$ from $U^{(j)}$ in a time that is polynomial in the number of 
qubits. It is currently not clear how to achieve this.

\section*{Acknowledgments}
I'm grateful to Anargyros Papageorgiou for very helpful discussions and comments.

\end{document}